\documentclass[10pt,journal,a4paper]{IEEEtran} 

\usepackage{amssymb}
\usepackage{amsmath}
\usepackage{bm}
\usepackage{amsthm}
\usepackage{graphicx}
\usepackage{subfigure}
\usepackage{float}
\usepackage{cite}
\usepackage{footnote}
\usepackage{threeparttable}
\makesavenoteenv{tabular}
\makesavenoteenv{table}
\usepackage{enumerate}
\usepackage{enumitem}
\usepackage{multirow}
\usepackage{amsfonts}
\usepackage{makecell} 
\usepackage{url}
\usepackage{amsthm}
\usepackage{color, soul}
\usepackage{cite}
\newtheorem{remark}{\bf~~Remark}

\begin{document}
	
\title{{RIS-based IMT-2030 Testbed for MmWave Multi-stream Ultra-massive MIMO Communications}}
\author{
{Shuhao Zeng},  
{Boya Di},
{Hongliang Zhang},
{Jiahao Gao},
{Shaohua Yue},\\
{Xinyuan Hu},
{Rui Fu},
{Jiaqi Zhou},
{Xu Liu},
{Haobo Zhang},
{Yuhan Wang},\\
{Shaohui Sun},
{Haichao Qin},
{Xin Su},
{Mengjun Wang},
{and Lingyang Song}


\thanks{S. Zeng, B. Di (corresponding author), H. Zhang, J. Gao, S. Yue, X. Hu, X. Liu, H. Zhang, Y. Wang, and L. Song are with Peking University.}


\thanks{R. Fu, and J. Zhou are with Xiangtan University.}

\thanks{S. Sun, H. Qin, X. Su, and M. Wang are with State Key Laboratory of Wireless Mobile Communications Technology (CATT) Beijing, China.}
}

\maketitle
\vspace{-3mm}
\begin{abstract}
As one enabling technique of the future sixth generation~(6G) network, ultra-massive multiple-input-multiple-output~(MIMO) can support high-speed data transmissions and cell coverage extension. However, it is hard to realize the ultra-massive MIMO via traditional phased arrays due to unacceptable power consumption. To address this issue, reconfigurable intelligent surface-based~(RIS-based) antennas are an energy-efficient enabler of the ultra-massive MIMO, since they are free of energy-hungry phase shifters. In this article, we report the performances of the RIS-enabled ultra-massive MIMO via a project called \emph{Verification of MmWave Multi-stream Transmissions Enabled by RIS-based Ultra-massive MIMO for 6G}~(V4M), which was proposed to promote the evolution towards IMT-2030. In the V4M project, we manufacture RIS-based antennas with $1024$ one-bit elements working at $26$ GHz, based on which an mmWave dual-stream ultra-massive MIMO prototype is implemented for the first time. To approach practical settings, the Tx and Rx of the prototype are implemented by one commercial new radio base station and one off-the-shelf user equipment, respectively. The measured data rate of the dual-stream prototype approaches the theoretical peak rate. Our contributions to the V4M project are also discussed by presenting technological challenges and corresponding solutions.


\end{abstract}
\vspace{-3mm}
\section{Introduction}

The future sixth generation~(6G) cellular network is required to achieve higher data rate and ubiquitous connectivity for supporting emerging applications such as virtual reality and augmented reality~\cite{saad2019vision}. To fulfill this vision, ultra-massive multiple-input-multiple-output~(MIMO) is a key enabler, which can provide significant multiplexing gain due to its massive elements and large radiation aperture. However, it is non-trivial to realize the ultra-massive MIMO using traditional phased arrays. This is because the phased array requires hundreds of energy-intensive phase shifters, which can lead to unacceptable power consumption.

Recently, reconfigurable intelligent surface-based~(RIS-based) antennas provide a more energy efficient solution for the implementation of the ultra-massive MIMO. To be specific, the RIS is an ultra-thin surface consisting of multiple sub-wavelength scattering elements~\cite{zeng2020reconfigurable}. Each element can reflect the incident signals from transmit antennas (often referred to as feeds) and apply adjustable phase shifts. By controlling the biased voltages over the diodes on the RIS elements, the phase shifts induced by the elements can be reconfigured to shape reflected beams into a desired form. Different from the phased arrays, the RIS-based antennas do not contain any phase shifters, and the power consumption of each RIS element is ultra-low~\cite{yin2020single}. Thus, the RIS-based antennas consume less power than the phased arrays. 

{Apart from the RIS-based antenna, the development of the RIS also introduces another wireless technology called holographic MIMO~\cite{huang2020holographic}, which is deployed between the BS and the user equipments~(UEs) to shape the wireless propagation environments into a desirable form. Unlike the holographic MIMO, the RIS-based antenna is usually regarded as an energy efficient solution to the ultra-massive MIMO, where the RIS is deployed at the BS for beamforming.}

Existing research works on the RIS-based antenna mainly focus on the theoretical analysis~\cite{zeng2022reconfigurable,zeng2022multiuser} and experimental verifications~\cite{dai2020reconfigurable} of wireless communications enabled by the RIS-based antennas. The authors in~\cite{zeng2022reconfigurable} analyze the capacity and power consumption of a single-user network enabled by the RIS-based antenna. In~\cite{zeng2022multiuser}, a BS utilizes one RIS-based antenna to serve multiple UEs simultaneously, where the size of the antenna is optimized to maximize the energy efficiency of the system. To experimentally verify the power-saving capability of RIS-based antennas, the authors in~\cite{dai2020reconfigurable} implement an RIS-based antenna with $256$ elements working at $2.3$~GHz, based on which a single-stream point-to-point communication testbed is deployed and evaluated. \textit{However}, due to the limited number of RIS elements working in the sub-6G band, the experimental results in~\cite{dai2020reconfigurable} cannot characterize the performance of the RIS-based ultra-massive MIMO in the typical mmWave band. Besides, multi-stream transmissions have not been considered, which are important for data rate enhancement.

In this article, we present the performances of the RIS-based ultra-massive MIMO via a research project called \emph{Verification of MmWave Multi-stream Transmissions Enabled by RIS-based Ultra-massive MIMO for 6G}~(V4M). Notably, V4M was launched in order to promote the evolution towards IMT-2030, i.e., international mobile telecommunication for 2030 and beyond, which is expected to achieve higher energy efficiency and peak data rate~\cite{IMT2030}. Our contributions to the V4M project are summarized as below, 
\begin{enumerate}[itemindent=0em, label=$\bullet$]
	\item {\textbf{Implementation of mmWave RIS-based antennas}: To support the ultra-massive MIMO, we design a one-bit RIS element through a circuit-based method and deploy the feed by combining numerical optimizations and full-wave simulations, based on which a RIS-based antenna with $1024$ elements working at $26$ GHz is manufactured.}
	\item \textbf{Prototyping of RIS-based multi-stream ultra-massive MIMO systems}: We, for the first time, deploy a dual-stream RIS-based ultra-massive MIMO system using the implemented mmWave RIS-based antennas. To approach practical settings, the Tx and Rx of the system are realized by utilizing a commercial new radio~(NR) BS and an off-the-shelf user equipment~(UE), respectively. 
	\item \textbf{Experimental evaluations}: We evaluate the prototypes in an indoor environment to measure typical parameters. {Experimental results show that the RIS-based ultra-massive MIMO is a promising solution to achieve the high data rate and low power consumption requirements of 6G. Besides, the measurement results are made publicly available, which can boost further study on RIS.}
\end{enumerate}



The rest of this paper is organized as follows. First, we provide basic information of the V4M project. Then, the implementation of the RIS-based antennas and corresponding wireless communication testbeds are introduced. Following that, the experimental results of the implemented prototypes are illustrated. Potential extensions to the V4M project are also discussed. Finally, conclusions are drawn in Section~\ref{sec_conclusion}.

  
\section{Verification of MmWave Multi-stream Transmissions Enabled by RIS-based Massive MIMO for 6G~(V4M)}
\label{sec:principle} 
\subsection{Project Objective}
\label{prj_obj}
The V4M project aims to accurately characterize the performances of the RIS-based ultra-massive MIMO in one typical application scenario, i.e., mmWave multi-stream transmissions enabled by RIS-based ultra-massive MIMO. To this end, we need to implement large scale RISs working in the millimeter-wave band, based on which a multi-stream communication testbed should be deployed. Besides, to capture the features of practical RIS-based communication systems, the transceivers in the multi-stream testbed should meet the technical specifications developed by 3GPP.

%

\subsection{Research Roadmap}
To evaluate the performances of RIS-based wireless communications, we take three steps from hardware module designs to system evaluations:
\begin{itemize}
	\item \textbf{Step 1}: Implementation of RIS. { The first challenge is how to design a RIS element with a desired reflection amplitude and phase by optimizing the structure parameters of the element, which is non-trivial since the RIS element contains many structure parameters to be optimized, and these parameters are coupled. In addition, it is challenging to optimize the deployment of the feed in order to maximize the gain of the RIS-based antenna, since it is time-consuming to evaluate possible deployments through full-wave simulations.}


	\item \textbf{Step 2}: Prototyping RIS-based wireless communication systems. To characterize the performances of the RIS-based ultra-massive MIMO in the typical application scenario mentioned in Section~\ref{prj_obj}, we need to realize an mmWave multi-stream communication testbed using RIS-based antennas of large aperture. 
	\item \textbf{Step 3}: Experimental evaluation. The implemented prototypes should be deployed in typical experimental setups, where their key performance metrics, such as error vector magnitude, data rate, etc., will be measured.
\end{itemize}

\begin{figure*}[!t]
	\centering
	\includegraphics[width=0.80\textwidth]{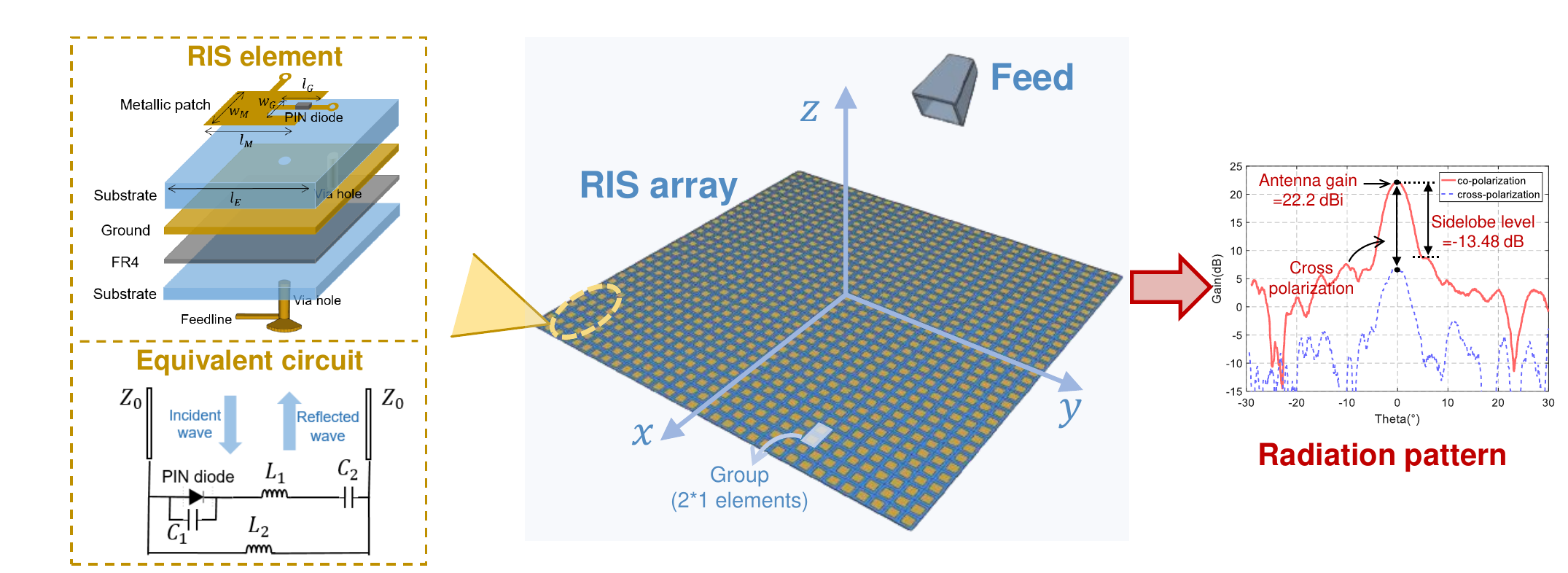}
	\caption{Implementation of RIS.}
	\label{implementation_RIS}
\end{figure*}

\section{Implementation of RIS-based antenna}
In this part, we describe the implementation of a one-bit linear-polarized RIS-based antenna working at $26$~GHz. 
\subsection{Structure of RIS Elements}
To realize a one-bit RIS element, we adopt the element structure shown in Fig.~\ref{implementation_RIS}. To be specific, on the top of the element is a positive-intrinsic-negative~(PIN) diode and a metallic patch with a groove, which are supported by a dielectric substrate of material RO4350B. On the other side of the substrate is a metallic ground for avoiding signal leakage so as to enhance the strength of reflection signals. The ground is connected to the metallic patch through a via hole and the PIN diode, which can be switched between \emph{ON} and \emph{OFF} states according to predetermined biased voltages. We would like to point out that the state of the RIS element is determined by the state of the PIN diode, and thus the RIS element can be configured to two states. To provide the required biased voltages, a feedline connecting with the metallic patch through a via hole is deployed at the bottom of the RIS element. We would like to point out that a microstrip radial stub is attached to the feedline to choke RF signals so as to reduce insertion loss. To support the feedline and the radial stub, another dielectric substrate of material RO4350B is utilized, which is attached to the ground using FR4. 


\subsection{Optimization of Structure Parameters of RIS Elements}
To improve the beamforming gain provided by the RIS, we need to enhance the reflection amplitude of the RIS element at the working frequency of $26$~GHz. Besides, it is desirable to reach a reflection phase difference of $180^{\circ}$ between the two states of the element. {To achieve the desired reflection response, the structure parameters of the RIS element should be optimized, which is challenging since the structure of the RIS element is complex with numerous parameters to be optimized, as shown in Fig.~\ref{implementation_RIS}, and these structure parameters are coupled. To cope with this issue, we propose a circuit-based method, where the equivalent circuit of the RIS element is modeled first, as shown in Fig.~\ref{implementation_RIS}. Based on this circuit model, the key structure parameters that influence the reflection performances can be determined, i.e., the period $l_E$ of the element, the size $l_M\times w_M$ of the metallic patch, and the size $l_G\times w_G$ of the groove on the metallic patch. Therefore, the number of structure parameters to be optimized can be reduced. To cope with the coupling among these structure parameters, we adopt an alternative optimization based method, where the structure parameters are optimized one by one through parameter sweeping by performing full-wave simulations. The key structure parameters are optimized as $l_E=5$~mm, $l_M\times w_M=2.68\times1.92$~mm$^2$, and $l_G\times w_G=0.74\times 0.86$~mm$^2$. Besides, the corresponding reflection amplitude is larger than $0.85$ for both states of the RIS element, and the reflection phase difference between the two states is $180.8^{\circ}$, thus verifying the proposed circuit-based method.}



\subsection{Design of RIS Array}
Based on the designed RIS element, we implement a $32\times 32$ RIS array with a size of $16\times 16$~cm$^2$. To provide biased voltages to these RIS elements, a feeding network is deployed on the back of the RIS to connect the feedline of each element to an RIS controller. We would like to point out that it is challenging to provide an independent biased voltage to each of the $1024$ elements, since it requires a highly complex RIS controller and feeding network. For the sake of design complexity, we divide the RIS elements into several groups, where the elements within the same group share the same biased voltage. Although combining more RIS elements into one group can further reduce design complexity, it can also degrade the beam steering capability of the RIS. Therefore, to achieve a tradeoff between the design complexity and beam steering capability, we set each group to contain two RIS elements, as shown in Fig.~\ref{implementation_RIS}, where a beam-scanning range of $[-60^{\circ},60^{\circ}]$ in the horizontal plane and $[-10^{\circ},10^{\circ}]$ in the vertical plane can be achieved under this setting. The RIS controller is implemented by field-programmable gate arrays~(FPGAs) of model Cyclone EP4CE10F17C8, which converts received configurations from a host into corresponding biased voltages so as to shape the beam of the RIS into a desirable form.

\subsection{Deployment of Feed}
\label{subsec:deployment_of_feed}
In general, an RIS-based antenna consists of one RIS illuminated by one transmit antenna, which is often referred to as \emph{feed}~\cite{zeng2022reconfigurable}. Therefore, to implement an RIS-based antenna, a horn antenna of model LB-34-10-C-KF is selected as the feed, as shown in Fig.~\ref{implementation_RIS}.{The deployment of the feed should be optimized since it has an influence on the gain of the RIS-based antenna~\cite{jiang2014folded}, which, however, is challenging and complex since the feed can be deployed in various possible locations and it takes a long time, i.e., nearly three hours, to evaluate one possible location through full-wave simulations. To cope with this issue, we design an efficient feed deployment scheme by combining numerical optimizations and full-wave simulations. To be specific, we first model the antenna gain of the RIS-based antenna with respect to the deployment of the feed, based on which the deployment of the feed is optimized through numerical optimization methods. Although the numerial optimization is time-saving, the derived deployment is not optimal since the antenna array gain model is just an approximation. Note that we can acquire the accurate antenna array gain through full-wave simulations. Therefore, we further evaluate several possible locations around the location derived in the first step through full-wave simulations in order to optimize the deployment of the feed. Through the proposed scheme, the derived optimal coordinate of the feed is $(-82,0,150)$~mm.} To evaluate the performance of the designed RIS-based antenna, we measure its radiation pattern in a microwave anechoic chamber, which is recorded in Fig.~\ref{implementation_RIS}. Experimental results demonstrate that at the working frequency of $26$~GHz, the designed RIS-based antenna has a high antenna gain of $22.2$ dBi, a low sidelobe level~(SLL) of $-13.48$ dB, and a cross polarization of below $-15$~dB.

\section{Prototyping of RIS-based wireless communication system}
\begin{figure*}[!t]
	\centering
		\includegraphics[width=0.80\textwidth]{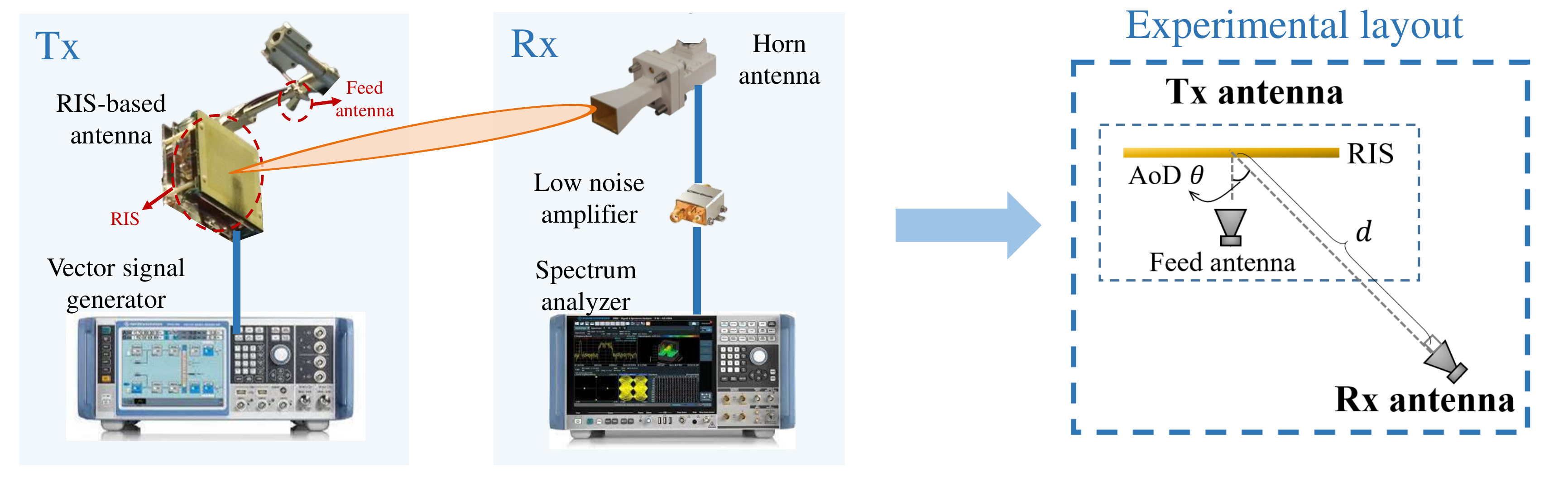}
	\caption{Hardware modules of the implemented single-stream RIS-enabled prototype and corresponding experimental layouts.}
	\label{hardware_single_stream_antenna}
\end{figure*}
In this section, a \emph{step-by-step approach} is adopted to develop the RIS-based wireless communication testbed. A point-to-point single-stream communication prototype is developed first, which is then extended to a more general dual-stream system.

\subsection{Single-stream system}
\label{sub_sub_section_single_stream}
As shown in Fig.~\ref{hardware_single_stream_antenna}, the single-stream communication prototype includes the following components:
\begin{itemize}
	\item \textbf{Transmitter}: The transmitter~(Tx) is constituted by one vector single generator and the designed RIS-based antenna. The type of the vector signal generator is SMW200A, which implements the baseband processing and up-conversion. The radio-frequency~(RF) signals output by the vector signal generator are radiated out through the RIS-based antenna. To direct the transmitted signals towards the receiver, we design the states of the RIS elements by jointly considering the location of the feed, the geometry of the RIS, and the target beam direction~\cite{wan2016field}.
	\item \textbf{Receiver}: The receiver~(Rx) includes one horn antenna, one low noise amplifier~(LNA), and one signal and spectrum analyzer. The RF signal received by the horn antenna is first amplified by the LNA, and is then sent to the signal and spectrum analyzer for spectrum and transmission error characterization. The part number of the signal and spectrum analyzer is FSW, and the gains of the antenna and the LNA are $22$~dBi and $30$~dB, respectively.
\end{itemize}
\subsection{Dual-stream system}


\begin{figure*}[!t]
	\centering
	\includegraphics[width=0.75\textwidth]{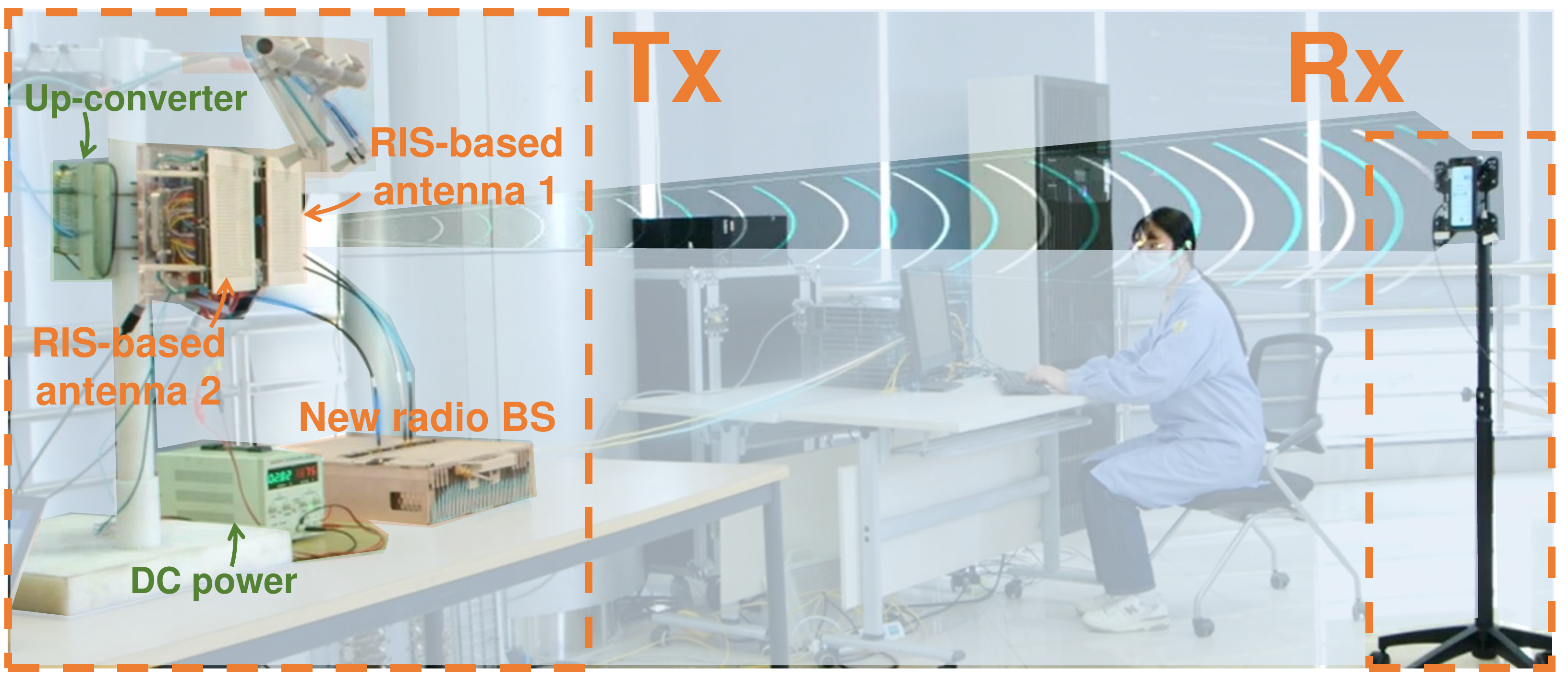}
	\caption{Hardware modules of the dual-stream RIS-enabled communication prototype and corresponding experimental layouts.}
	\label{experiment_setup_single_stream_antenna}
\end{figure*}

In the dual-stream system, one Tx simultaneously sends two data streams to one Rx by utilizing the polarizations of RIS-based antennas, as shown in Fig.~\ref{experiment_setup_single_stream_antenna}. The hardware architectures of the Tx and the Rx are shown below.
\begin{itemize}
	\item \textbf{Transmitter}: The Tx consists of one commercial 5G NR BS, one up-converter, and two designed RIS-based antennas with orthogonal polarizations\footnote{By simplicity rotating each RIS element and the feed by $90^{\circ}$ around their normals, we can obtain another linearly polarized RIS-based antenna whose polarization is orthogonal to that of the original antenna}. The BS first performs baseband processing and carrier modulation over the two data streams. The output two intermediate~(IF) signals are shifted to the mmWave band through the up-converter, which are then transmitted by the two RIS-based antennas, respectively. The orthogonality of polarization between the two RIS-based antennas, i.e., horizontal and vertical polarizations, respectively, is utilized to suppress inter-stream interference and support the dual-stream transmissions. 
	
	\item \textbf{Receiver}: The Rx is an off-the-shelf UE powered by MediaTek's 5G baseband modem of type M80. A dual-polarized antenna is used by the Rx to receive the two data streams. Due to the low cross polarization of the RIS-based antennas as recorded in Section~\ref{subsec:deployment_of_feed}, the inter-stream interference can be effectively canceled, which enables high-speed dual-stream transmissions between the Tx and the Rx.  
\end{itemize}
 
\section{Experimental Evaluation}
\subsection{Single-stream system}
\label{sub_sub_single}

We characterize the performances of the implemented RIS-based single-stream system by measuring the following key metrics:
\begin{itemize}
	\item \textit{Error vector magnitude}~(EVM): 
	EVM is defined as the normalized root mean square amplitude of the error between received actual symbols and ideal constellation points. {Here, we evaluate the EVM since it is a widely used system-level performance metric that characterizes the reliability of wireless communications.} According to 3GPP, the EVM of the single-stream prototype should not exceed $8\%$ in order to guarantee successful data transmissions~\cite{3gpp.38.104}, given that 64 quadrature amplitude modulation~(64QAM) is adopted in the system as we will mention later.
	\item \textit{Adjacent channel leakage ratio}~(ACLR): ACLR is defined as the ratio between the power $P_d$ within designated channels and the power $P_a$ within adjacent channels, i.e., $\frac{P_a}{P_d}$. It is a measurement of interference to the adjacent channels, which is required to be lower than $-28$~dBc according to 3GPP~\cite{3gpp.38.104}.
\end{itemize}

\begin{figure*}[!t]
	\centering
	\includegraphics[width=0.80\textwidth]{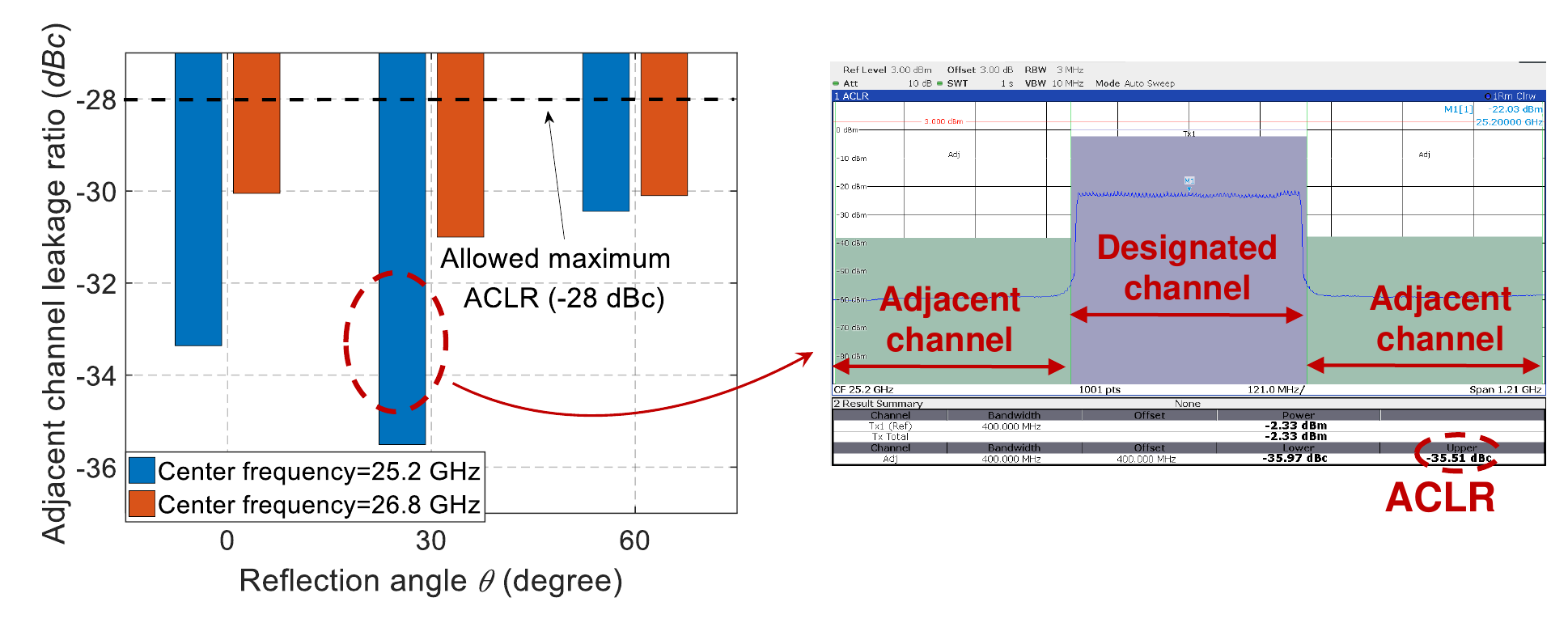}
	\caption{Adjacent channel leakage ratio~(ACLR) vs. AoD $\theta$, with Tx-Rx distance $d=4$~m.}
	\label{ACLR_vs_theta}
\end{figure*}

The single-stream communication prototype is deployed in a typical meeting room. The Tx generates a 64QAM signal with a center frequency of $25.2$~GHz, a bandwidth of $400$~MHz and a power of $1$~dBm. {For the simplicity of discussions, we define $\theta$ as the angle of departure~(AoD) from the RIS-based antenna to the Rx antenna, and use $d$ to represent the distance between the Rx antenna and the center of the RIS array, as shown in Fig.~\ref{hardware_single_stream_antenna}. 
	

%
%


\begin{figure*}[!t]
	\centering
	\includegraphics[width=0.80\textwidth]{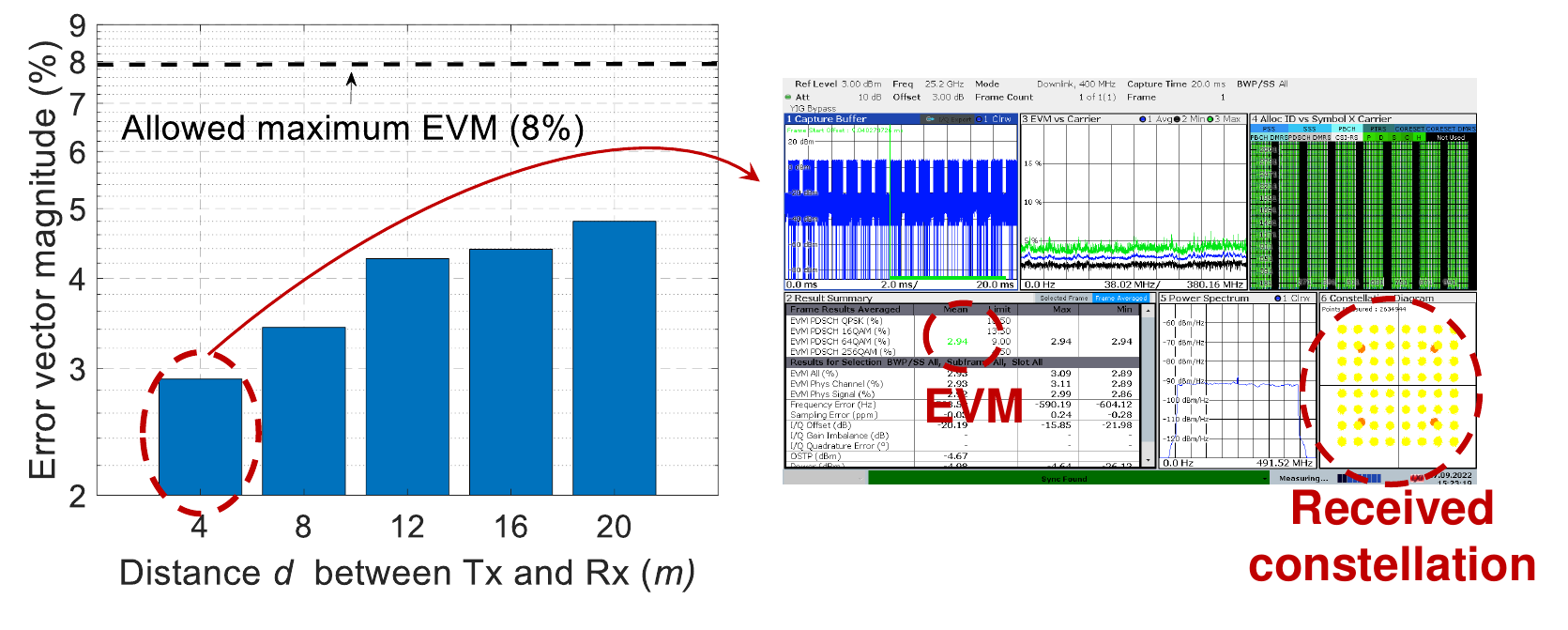}
	\caption{Error vector magnitude~(EVM) vs. Tx-Rx distance $d$, with AoD $\theta=0^{\circ}$.}
	\label{EVM_vs_d}
\end{figure*}

Fig.~\ref{EVM_vs_d} depicts the EVM versus the distance $d$ between the Tx and the Rx. We can find that as the Tx moves away from the Rx, the EVM becomes larger, which indicates higher bit error rate. This is because the pathloss from the Tx to the Rx becomes more severe, leading to degraded received signal-to-noise ratio. However, we can find that even when the Tx and Rx are separated at a distance of $20$~m, the EVM still satisfies the 3GPP requirement, i.e., $\le 8\%$.

In Fig.~\ref{ACLR_vs_theta}, we plot ACLR versus AoD $\theta$. The bandwidth of the adjacent channels are set as $400$~MHz. Besides, we consider the case when the center frequency of the designated channel is $25.2$~GHz and $26.8$~GHz, respectively. We can find that under the two center frequencies, the ACLR always satisfies the 3GPP requirement, i.e., $<-28$~dBc, when the AoD ranges from $0^{\circ}$ to $60^{\circ}$. By combining Fig.~\ref{EVM_vs_d} and Fig.~\ref{ACLR_vs_theta}, we can conclude that \emph{the RIS-based ultra-massive MIMO can support the single-stream communications.}
}

\subsection{Dual-stream system}
\label{sub_sub_dual}
The dual-stream communication prototype is deployed in the same meeting room as the single-stream communication prototype, as shown in Fig.~\ref{experiment_setup_single_stream_antenna}. The data is transmitted using the NR Physical Downlink Shared Channel~(PDSCH), where 64 QAM is adopted. The DDDSU frame structure is adopted to organize transmitted symbols, where the special ``S" slot is configured as $10:2:2$. Besides, $4$ component carriers~(CCs) are aggregated and allocated to the UE, where each CC has a bandwidth of $200$~MHz. The center frequency of the transmitted signals are set as $26.6$~GHz. We deploy the Rx directly in front of the Tx, where the distance between the Tx and the Rx is $3$~m.

Table~\ref{state} records the measured performances of the dual-stream communication testbeds, where H-polarized and V-polarized represent horizontal and vertical polarizations, respectively. We can find that since the sidelobe is significantly suppressed, as indicated by the low SLL, the gains of both RIS-based antennas are high. Note that the cross-polarization components generated by the two RIS-based antennas are weak. Therefore, the received signal-to-interference-and-noise-ratio~(SINR) is high, and thus a high data rate of $5$~Gbps can be achieved, which approaches the theoretical peak data rate given the same bandwidth, the same number of data streams, and the same modulation order~\cite{3gpp.38.306}. These measured results collaboratively verify that the RIS-based ultra-massive MIMO can support multi-stream communications in the mmWave band. Besides, we can also find that compared with the conventional phased arrays presented in~\cite{kibaroglu201864}, the implemented RIS-based antennas can reduce the power consumption by $38\%$ while achieving similar antenna gains. This indicates that the RIS-based antennas are an energy efficient enabler of the ultra-massive MIMO. 

\begin{table*}[!t]
	\renewcommand\arraystretch{1.2}
	\caption{Measured performances of the dual-stream RIS-enabled communication testbeds (\text{working frequency: $26.6$~GHz}).}
	\centering
	\small
	\begin{tabular}
		{| p{0.1\textwidth}<{\centering} | p{0.08\textwidth}<{\centering} | p{0.08\textwidth}<{\centering} |  p{0.13\textwidth}<{\centering}| p{0.25\textwidth}<{\centering} |p{0.2\textwidth}<{\centering}|}
		\Xhline{1.pt}
		\textbf{RIS-based antenna} & \textbf{Antenna gain} (dBi) & \textbf{Sidelobe level} (dB) & \textbf{Cross polarization} (dB) & \textbf{Data rate (two streams)} & \textbf{Power consumption}\\
		\hline
		H-polarized & 22.01 & -7.43 & -15.19 &\multirow{2}*{\makecell{Measured data rate: 5~Gbps \\ \footnotesize{(Theoretical peak rate\footnotemark[3]: 5.17~Gbps~\cite{3gpp.38.306})}}} 
		&\multirow{2}*{\makecell{RIS-based antennas: 15.8~W\\\footnotesize{(Phased array: 25.6~W~\cite{kibaroglu201864})}}}\\
		\cline{1-4}
		V-polarized & 22.11 & -13.55 & -10.16 & &\\
		\Xhline{1.pt}
	\end{tabular}
	\label{state}
\end{table*}
\begin{remark}
	The prototype parameters, experimental setups and measured results related with the V4M project can also serve as useful references for the numerical simulations on the RIS-based wireless communications. Due to the space limit, we only present a part of the aforementioned information here. For interested readers, full information can be found in the following website: \textit{https://github.com/Wonderzeng/Verification-of-RIS-based-wireless-communications-for-IMT-2030}. 
	
\end{remark}

{From the actual testing of the prototype, we have found the following defects and limitations of the RIS, i.e.,}
\begin{remark}
	{The reflection coefficients of the RIS elements depend on the angle of incidence, which, however, is not considered in most existing beamforming algorithms, and thus the beam can deviate from intended receivers. To improve the system performance, the angle of incidence and the corresponding reflection coefficient should be estimated for each RIS element so as to direct the beam towards the intended receivers. 
		
	In addition, when we synthesize wide-lobe beams from narrow beams, the low-resolution phase shift of the RIS elements can lead to undesired fluctuations in the main lobe of the generated beams. Such fluctuation can degrade the performance of traditional hierarchical beam training, since the beam search space gradually narrows layer by layer, and the misjudgment of beams will cause the beam search to ``go away" in the wrong direction, which cannot be corrected. To guarantee the performance of beam discrimination, we should enlarge the beam search space to evaluate more codewords if none of the codewords within current beam search space perform well.}
\end{remark}

\begin{remark}
	{To address the reproducibility of the experiment, we remeasure the radiation performances of the implemented RIS-based antenna again in the same microwave chamber. The acquired radiation patterns from the two measurements coincide, which verifies the reproducibility of the experiment.}
\end{remark}

\section{Potential Extensions and Research Challenges}
Previously, we have introduced in detail the contents of the V4M project. In this section, we briefly discuss some potential extensions to this project and related research challenges.  
\footnotetext[3]{The theoretical peak data rate is calculated given the same bandwidth, the same number of data streams, and the same modulation order as the dual-stream prototype.}
\subsection{RIS-assisted High-mobility Wireless Systems}
As a key performance indicator, the future 6G network is required to provide services to high-mobility terminals which can travel at a speed up to $1000$~km/h~\cite{jiang2021road}. Therefore, the RIS-based antenna, which is a potential technique of the 6G network, is also expected to serve users with high mobility. 

However, in the high-mobility scenario, the estimation of the RIS-based channels can be more challenging due to Doppler frequency shift. In fact, the existence of significant Doppler frequency shift can lead to short channel coherent time. Therefore, channel estimation methods with low computation complexity and short pilot overhead are required to ensure the timeliness of the estimated channel state information and to guarantee effective data transmissions. In this context, compressive-sensing-based channel estimation methods may be exploited to meet the stringent requirements.
\vspace{-0.1cm}

\subsection{Beam Training with Unknown CSI for RIS}
Traditional optimization-based beamforming schemes for the RIS-based antennas require accurate channel state information~(CSI) of RIS-based communications, which, however, is hard to obtain due to the passive characteristic of the RISs and the massive number of RIS elements. To cope with this challenge, beam training methods serve as a promising solution since they do not require any CSI. 

It is, however, non-trivial to design the codebook at the RIS based antennas for the beam training methods. One challenge lies in how to optimize the size of the codebook to achieve a tradeoff between the beam accuracy and pilot overhead. Besides, since the beams transmitted by the RIS-based antennas are closely related to both the digital beamformer at the BS and the phase shifts of the RIS elements, the codebook at the RIS and that at the BS should be jointly designed. {We would like to point out that although the codebook-based solutions for RIS has been studied in the literature, as summarized in~\cite{an2022codebook}, these research efforts do not apply to the RIS-based ultra-massive MIMO considered in this paper. This is because in these works, the RIS is deployed far from the BS and acts as a relay, while in this paper, the RIS is much closer to the BS antenna (i.e., feed), and thus the BS antenna cannot be assumed to locate in the far field of the RIS.}
\vspace{-0.1cm}

{
\subsection{RIS-based near-field communication}
Due to the low power consumption of the RIS-based ultra-massive MIMO, it is possible to further extend the scale of the antenna array to fullfil the high spectrum efficiency requirement of 6G. However, the increased radiation aperture of the antenna array leads to significantly enlarged Fresnel region, i.e., radiating near-field region, indicating that the users are located within the near-field region of the antenna array. As an example, for an antenna array with a size of $0.5\times 0.5$~m$^2$ working at $30$~GHz, the range of the near field is $100$~m, which is comparable to the cell radius. 

Different from traditional far-field communications, the EM waves in the RIS-based near-field communications can be better characterized by spherical waves than plane waves, which brings a new challenge to beamforming design. To be specific, since the far-field array response vector only depends on the angle of arrival/departure~(AoA/AoD) while the near-field array response vector is influenced by both the angle and the distance due to spherical curvature, the conventional far-field beamforming algorithm should be revised to take such distance-dependent characteristic into account before it can be applied to the RIS-based near-field communications.
\vspace{-0.1cm}

\subsection{RIS-based integrated sensing and communication}
Integrated sensing and communication~(ISAC) has attracted much attention due its capability of alleviating spectrum congestion. To provide high spatial diversity to the ISAC system, the RIS-based ultra-massive MIMO is a promising solution with low power consumption, where the RIS-based antenna is utilized as the BS antenna for beamforming.

However, the integration of the RIS and the ISAC brings new challenges to the configuration of the RIS. First, to support high communication rate and high-resolution parameter estimation, wideband waveforms should be adopted in the RIS-based ISAC system. However, the RIS elements are in general frequency-selective, i.e., showing different electromagnetic responses at different frequencies, which makes the RIS-based beamforming more challenging. In addition, it is non-trivial to achieve a tradeoff between the sensing and communications by configuring the RIS.}
\vspace{-0.25cm}
\section{Conclusions}
\vspace{-0.05cm}
\label{sec_conclusion}
In this paper, we have designed and manufactured RIS-based antennas with $1024$ one-bit elements working at $26$~GHz. Based on the implemented RIS-based antennas, an mmWave single-stream ultra-massive MIMO prototype has been deployed first, which has then been extended to a dual-stream communication testbed. To approach practical settings, the Tx and Rx of the dual-stream prototype have been implemented by one commercial NR BS and one off-the-shelf UE, respectively. Both prototypes have been evaluated in an indoor environment to measure typical performance metrics such as EVM, data rate, etc. According to experimental results, the RIS-enabled dual-stream testbed can achieve a data rate of $5$~Gbps, which approaches the theoretical peak rate. This shows that RIS-based ultra-massive MIMO can support multi-stream communications in the mmWave band.


%
%
%



\vspace{-2.5mm}
\begin{thebibliography}{20}	
\bibitem{saad2019vision}
W.~Saad, M.~Bennis, and M.~Chen, ``{A vision of 6G wireless systems: Applications, trends, technologies, and open research problems}," \emph{IEEE Netw.}, vol.~34, no.~3, pp.~134-142, May 2019.

\bibitem{zeng2020reconfigurable}
S.~Zeng, H.~Zhang, B.~Di, Z.~Han, and L.~Song, ``{Reconfigurable intelligent surface (RIS) assisted wireless coverage extension: RIS orientation and location optimization}," \emph{IEEE Commun. Lett.}, vol.~25, no.~1, pp.~269-273, Sep. 2020.

\bibitem{yin2020single}
J.~Yin, Q.~Wu, Q.~Lou, H.~Wang, Z.~Chen, and W.~Hong, ``{Single-beam 1 bit reflective metasurface using prephased unit cells for normally incident plane waves}," \emph{IEEE Trans. Antennas Propag.}, vol.~68, no.~7, pp.~5496-5504, Mar. 2020.

\bibitem{huang2020holographic}
C.~Huang, S.~Hu, G.~C. Alexandropoulos, A.~Zappone, C.~Yuen, R.~Zhang, M.~Di~Renzo, and M.~Debbah, ``{Holographic MIMO surfaces for 6G wireless networks: Opportunities, challenges, and trends}," \emph{IEEE Wireless Commun.}, vol.~27, no.~5, pp.~118-125, Oct. 2020.

\bibitem{zeng2022reconfigurable}
S.~Zeng, H.~Zhang, B.~Di, H.~Qin, X.~Su, and L.~Song, ``{Reconfigurable refractive surfaces: An energy-efficient way to holographic MIMO}," \emph{IEEE Commun. Lett.}, vol.~26, no.~10, pp.~2490-2494, Jul. 2022.

\bibitem{zeng2022multiuser}
S.~Zeng, H.~Zhang, B.~Di, and L.~Song, ``{Multi-user holographic MIMO systems: Reconfigurable refractive surface or phased array?}" in \emph{Proc. IEEE GLOBECOM}, Rio de Janeiro, BR, Dec. 2022.

\bibitem{dai2020reconfigurable}
L.~Dai, B.~Wang, M.~Wang, X.~Yang, J.~Tan, S.~Bi, S.~Xu, F.~Yang, Z.~Chen, M.~Di~Renzo \emph{et~al.}, ``{Reconfigurable intelligent surface-based wireless communications: Antenna design, prototyping, and experimental results}," \emph{IEEE access}, vol.~8, no.~1, pp.~45913-45923, Mar. 2020.

\bibitem{IMT2030}
{The Next G Alliance}, ``{Reply liasion statement to external organizations, development of IMT vision for 2030 and beyond}," 2021. [Online]. Available:
\url{https://www.nextgalliance.org/wp-content/uploads/2021/09/NextG_SG-2021-00085R000.pdf}

\bibitem{jiang2014folded}
M.~Jiang, W.~Hong, Y.~Zhang, S.~Yu, and H.~Zhou, ``{A folded reflectarray antenna with a planar SIW slot array antenna as the primary source}," \emph{IEEE Trans. Antennas Propag.}, vol.~62, no.~7, pp.~3575-3583, Apr. 2014.

\bibitem{wan2016field}
X.~Wan, M.~Qi, T.~Chen, and T.~Cui, ``Field-programmable beam reconfiguring based on digitally-controlled coding metasurface," \emph{Sci. Rep.}, vol.~6, no.~1, pp.~1-8, Feb. 2016.

\bibitem{3gpp.38.104}
{3GPP TS 38.104 version 17.7.0}, ``{Base Station (BS) radio transmission and reception (Release 17)}," Tech. Rep., 09-2022. [Online]. Available: \url{https://portal.3gpp.org/desktopmodules/Specifications/SpecificationDetails.aspx?specificationId=3202}

\bibitem{3gpp.38.306}
{3GPP TS 38.306 version 17.3.0}, ``{User equipment (UE) radio access capabilities (Release 17)}," Tech. Rep., 01-2023. [Online]. Available: \url{https://portal.3gpp.org/desktopmodules/Specifications/SpecificationDetails.aspx?specificationId=3193}

\bibitem{kibaroglu201864}
K.~Kibaroglu, M.~Sayginer, T.~Phelps, and G.~M. Rebeiz, ``{A 64-element 28-GHz phased-array transceiver with 52-dBm EIRP and 8--12-Gb/s 5G link at 300 meters without any calibration}," \emph{IEEE Trans. Microw. Theory Tech.}, vol.~66, no.~12, pp.~5796-5811, Dec. 2018.

\bibitem{jiang2021road}
W.~Jiang, B.~Han, M.~A. Habibi, and H.~D. Schotten, ``{The road towards 6G: A comprehensive survey}," \emph{IEEE Open J. Commun. Soc.}, vol.~2, pp.~334-366, Feb. 2021.

\bibitem{an2022codebook}
J.~An, C.~Xu, Q.~Wu, D.~W.~K. Ng, M.~Di~Renzo, C.~Yuen, and L.~Hanzo, ``Codebook-based solutions for reconfigurable intelligent surfaces and their open challenges," \emph{IEEE Wireless Communications}, to be published.
\end{thebibliography}
\vspace{-.5mm}

\begin{IEEEbiographynophoto}{Shuhao Zeng} [M’23] (shuhao.zeng@pku.edu.cn) is a postdoctoral associate in the Department of Electrical and Computer Engineering at Princeton University. He received the Ph.D. degree at the School of Electronics at Peking University in 2023. His current research interests include intelligent surfaces, ultra-massive MIMO and unmanned aerial vehicle networks.
\end{IEEEbiographynophoto}
\vspace{-1.2cm}
\begin{IEEEbiographynophoto}{Boya Di} [M’19] (boya.di@pku.edu.cn) is an assistant professor at the School of Electronics, Peking University, Beijing, China. She obtained her PhD degree from the Department of Electronics, Peking University in 2019. Prior to that, she received the B.S. degree in electronic engineering from Peking University in 2014. Her current research interests include reconfigurable intelligent surfaces, edge computing, vehicular networks, and aerial access networks. She received the best doctoral thesis award from China Education Society of Electronics in 2019. She is also the recipient of 2021 IEEE ComSoc Asia-Pacific Outstanding Paper Award. 
\end{IEEEbiographynophoto}
\vspace{-1.2cm}
\begin{IEEEbiographynophoto}{Hongliang Zhang} [M’19] (hongliang.zhang@pku.edu.cn) is an assistant professor at the School of Electronics, Peking University, Beijing, China. He received the Ph.D. degree at the School of Electrical Engineering and Computer Science at Peking University in 2019. His current research interest includes reconfigurable intelligent surfaces, aerial access networks, and game theory. He received the best doctoral thesis award from Chinese Institute of Electronics in 2019. He is the recipient of 2021 IEEE Comsoc Heinrich Hertz Award for Best Communications Letters.  
\end{IEEEbiographynophoto}
\vspace{-1.2cm}
\begin{IEEEbiographynophoto}{Jiahao Gao} (jiahao.gao@pku.edu.cn) received the B.S. degree in electronic information engineering from Beijing Institute of Technology, Beijing, China, in 2019. He is currently pursuing a Ph.D. degree with the Department of Electronics, Peking University, Beijing. His current research interests include reconfigurable intelligent surface and signal processing.
\end{IEEEbiographynophoto}
\vspace{-1.2cm}
\begin{IEEEbiographynophoto}{Shaohua Yue} (yueshaohua@pku.edu.cn) received his B.S. degree in electronic and information engineering from Peking University, China, in 2022, where he is currently pursuing a Ph.D. degree with the Department of Electronics. His current research interests include reconfigurable intelligent surfaces and massive MIMO. 
\end{IEEEbiographynophoto}
\vspace{-1.2cm}
\begin{IEEEbiographynophoto}{Xinyuan Hu} [S’22] (huxiny@pku.edu.cn) received her B.S. degree in electronic engineering from Peking University, China, in 2022, where she is currently pursuing a Ph.D. degree with the School of Electronics. Her current research interest is reconfigurable holographic surface.
\end{IEEEbiographynophoto}
\vspace{-1.2cm}
\begin{IEEEbiographynophoto}{Rui Fu} (furui1232021@163.com) received his B.S. degree in Internet of Things Engineering from Anhui University, China, in 2021. Currently, he is pursuing a MA.Eng degree with Hunan Institute of Advanced Sensing and Information Technology at Xiangtan University, China. His current research interest is reconfigurable intelligent surface.
\end{IEEEbiographynophoto}
\vspace{-1.2cm}
\begin{IEEEbiographynophoto}{Jiaqi Zhou} (jiaqi.zhou1108@outlook.com) received her B.S. degree in electronic engineering from Hunan Institute of Engineering, China, in 2021, where she is currently pursuing a MA.Eng degree with the Department of Electronics at Xiangtan University. Her current research interest is reconfigurable holographic surface. 
\end{IEEEbiographynophoto}
\vspace{-1.2cm}
\begin{IEEEbiographynophoto}{Xu Liu} [S'20] (xu.liu@pku.edu.cn) received the B.S. degree from the College of Engineering, Peking University, in 2020. He is currently pursuing the Ph.D. degree with the School of Electronics at Peking University. His current research interest is metamaterial-aided passive RF sensing techniques for the Internet of Things.
\end{IEEEbiographynophoto}
\vspace{-1.2cm}
\begin{IEEEbiographynophoto}{Haobo Zhang} [S’19] (haobo.zhang@pku.edu.cn) received his B.S. degree in electronic engineering from Peking University, China, in 2019, where he is currently pursuing the Ph.D. degree with the School of Electronics.
\end{IEEEbiographynophoto}
\vspace{-1.2cm}
\begin{IEEEbiographynophoto}{Yuhan Wang}  (yuhan.wang@stu.pku.edu.cn) received her B.E. degree in Electronic Information Science and Technology from Xiamen University, China, in 2022. She is currently pursuing a Ph.D. degree in communication and information systems at Peking University. Her current research interest is intelligent omni surface.
\end{IEEEbiographynophoto}
\vspace{-1.2cm}
\begin{IEEEbiographynophoto}{Shaohui Sun} (sunshaohui@catt.cn) received his Ph.D. from Xidian University, Xi’an, China, in 2003. Since January 2011, he has been the Chief Technical Officer with Datang Wireless Mobile Innovation Center of the Datang Telecom technology and industry group. In 2019, he joined Datang mobile communications equipment company Ltd. and currently serves as EVP of CICT Mobile Communications Technology Co. Ltd and one senior expert of the State Key Laboratory of Wireless Mobile Communications, CATT. Now, his research area of interest includes advanced technologies related to 5G/6G.
	
\end{IEEEbiographynophoto}
\vspace{-1.2cm}
\begin{IEEEbiographynophoto}{Haichao Qin} (qinhaichao@cictmobile.com) is currently with the Datang mobile communications equipment company Ltd., Beijing. His research interests include MIMO technology and intelligent surfaces.
\end{IEEEbiographynophoto}
\vspace{-1.2cm}
\begin{IEEEbiographynophoto}{Xin Su} (suxin@cictmobile.com) received the BS., MS. and Ph.D. degrees from Xidian University, Xi'an, China in 2000, 2003 and 2006 respectively. In 2007, he joined the System Lab, Telecommunication System Division, Samsung Electronics, Suwon, Korea. Since 2011, he has been with the Datang Wireless Mobile Innovation Center, CATT, Beijing, China. Currently, he is senior engineer of CICT Mobile Communications Technology Co. Ltd, Beijing, China. His research interests are multi-antenna systems. 
\end{IEEEbiographynophoto}
\vspace{-18cm}
\begin{IEEEbiographynophoto}{Mengjun Wang} (wangmengjun@catt.cn) received the M.S. degree in communication and information systems from the China Academy of Telecommunication Technology (CATT), Beijing, China. He is currently a Senior Engineer with the CATT. His research interests include mmWave mobile communications, MIMO technology, and heterogeneous wireless networks.
\end{IEEEbiographynophoto}
\vspace{-18cm}
\begin{IEEEbiographynophoto}{Lingyang Song} [F’19] (lingyang.song@pku.edu.cn) received his Ph.D. from the University of York, UK, in 2007, where he received the K. M. Stott Prize for excellent research. In May 2009, he joined the School of Electronics Engineering and Computer Science, Peking University, China, as a full Professor. His main research interests include cooperative and cognitive communications, physical layer security, and wireless ad hoc/sensor networks. He was a recipient of the IEEE Leonard G. Abraham Prize in 2016 and the IEEE Asia Pacific (AP) Young Researcher Award in 2012. He has been an IEEE Distinguished Lecturer since 2015.
\end{IEEEbiographynophoto}
\end{document}